\documentclass[12pt]{article}
\pdfoutput=1
\usepackage[latin1]{inputenc}

\usepackage{amsmath}
\usepackage{color}
\usepackage{amsfonts}
\usepackage{amssymb}
\usepackage{graphicx}
\usepackage{geometry}
\usepackage{amssymb,epsfig,subfigure}
\usepackage{hyperref}
\usepackage{comment}

\usepackage[numbers,sort&compress]{natbib}

\def\simge{%  ``greater than about'' symbol
    \mathrel{\rlap{\raise 0.511ex 
        \hbox{$>$}}{\lower 0.511ex \hbox{$\sim$}}}}

\def\simle{%  ``less than about'' symbol
    \mathrel{\rlap{\raise 0.511ex 
        \hbox{$<$}}{\lower 0.511ex \hbox{$\sim$}}}}

\makeatletter
\renewcommand\section{\@startsection {section}{1}{\z@}%
                                 {-3.5ex \@plus -1ex \@minus -.2ex}%nn
                                   {2.3ex \@plus.2ex}%
                                   {\normalfont\large\bfseries}}
\renewcommand\subsection{\@startsection{subsection}{2}{\z@}%
                                   {-3.25ex\@plus -1ex \@minus -.2ex}%
                                     {1.5ex \@plus .2ex}%
                                     {\normalfont\bfseries}}
\renewcommand\subsubsection{\@startsection{subsubsection}{3}{\z@}%
                                   {-3.25ex\@plus -1ex \@minus -.2ex}%
                                     {1.5ex \@plus .2ex}%
                                     {\normalfont\itshape}}
\makeatother

\def\pplogo{\vbox{\kern-\headheight\kern -29pt
\halign{##&##\hfil\cr&{\ppnumber}\cr\rule{0pt}{2.5ex}&\ppdate\cr}}}
\makeatletter
\def\ps@firstpage{\ps@empty \def\@oddhead{\hss\pplogo}%
  \let\@evenhead\@oddhead % in case an article starts on a left-hand page
}%      The only change in \maketitle is \thispagestyle{firstpage} instead of 
\def\maketitle{\par
 \begingroup
 \def\thefootnote{\fnsymbol{footnote}}
 \def\@makefnmark{\hbox{$^{\@thefnmark}$\hss}}
 \if@twocolumn
 \twocolumn[\@maketitle]
 \else \newpage
 \global\@topnum\z@ \@maketitle \fi\thispagestyle{firstpage}\@thanks
 \endgroup
 \setcounter{footnote}{0}
 \let\maketitle\relax
 \let\@maketitle\relax
 \gdef\@thanks{}\gdef\@author{}\gdef\@title{}\let\thanks\relax}
\makeatother

\numberwithin{equation}{section}

\newcommand*{\sgn}{\ensuremath{\mathrm{sgn}}}

\newcommand{\be}{\begin{eqnarray}}
\newcommand{\bea}{\begin{eqnarray}}
\newcommand{\ee}{\end{eqnarray}}
\newcommand{\eea}{\end{eqnarray}}
\newcommand\beq{\begin{eqnarray}}
\newcommand\eeq{\end{eqnarray}}

\def\be{\begin{eqnarray}}
\def\ee{\end{eqnarray}}
\def\ba#1\ea{\begin{align}#1\end{align}}
\def\bg#1\eg{\begin{gather}#1\end{gather}}
\def\bm#1\em{\begin{multline}#1\end{multline}}
\def\bmd#1\emd{\begin{multlined}#1\end{multlined}}

\def\({\left(}
\def\){\right)}
\def\[{\left[}
\def\]{\right]}

\begin{document}

\setcounter{page}0
\def\ppnumber{\vbox{\baselineskip14pt
%\hbox{hep-th/0000000}
}}
\def\ppdate{
%\footnotesize{SU/ITP-14/XX}
} \date{}

\author{Jing-Yuan Chen$^1$, Jun Ho Son$^1$, Chao Wang$^1$ and S. Raghu$^{1,2}$\\
[7mm] \\
{\normalsize \it $^{1}$Stanford Institute for Theoretical Physics, Stanford 
University, Stanford, CA 94305, USA} \\
{\normalsize \it $^2$SLAC National Accelerator Laboratory, 2575 Sand Hill Road, 
Menlo Park, CA 94025, USA }\\
}

\bigskip
\title{\bf Exact Boson-Fermion Duality on a 3D Euclidean Lattice\vskip 0.5cm}
\maketitle

\begin{abstract}
The idea of statistical transmutation plays a crucial role in descriptions of the fractional quantum Hall effect. However, a recently conjectured duality between a critical boson and a massless 2-component Dirac fermion extends this notion to gapless systems. This duality sheds light on highly non-trivial problems such as the half-filled Landau level, the superconductor-insulator transition, and surface states of strongly coupled topological insulators. Although this boson-fermion duality has undergone many consistency checks, it has remained unproven. We describe the duality in a non-perturbative fashion using an exact UV mapping of partition functions on a 3D Euclidean lattice.
%Our approach is purely analytic and has the advantage of being simple and exact. 
\end{abstract}
\bigskip

\newpage

%\tableofcontents

\vskip 1cm

%%%%%%%%%%%%%%%%%%%%%%%%%%%%%%%%%%%%%%%%%%%%%%%%%%%%%%%%%%%%%%%%%
%%%%%%%%%%%%%%%%%%%%%%%%%%%%%%%%%%%%%%%%%%%%%%%%%%%%%%%%%%%%%%%%%
%%%%%%%%%%%%%%%%%%%%%%%%%%%%%%%%%%%%%%%%%%%%%%%%%%%%%%%%%%%%%%%%%
%%%%%%%%%%%%%%%%%%%%%%%%%%%%%%%%%%%%%%%%%%%%%%%%%%%%%%%%%%%%%%%%%
\section{Introduction}\label{sec:intro}

The idea of duality underlies some of the most fascinating aspects of quantum statistical mechanics. A system that appears strongly coupled and nearly intractable may be dual to a weakly interacting system. Duality thus enables us to solve some highly non-trivial problems in theoretical physics. In $D = 2 + 1$ dimensions, a class of duality mappings known as \emph{particle-vortex duality} follow from the fact that conserved currents can be expressed in terms of dual electric and magnetic fluxes. The simplest such example is the \emph{boson-vortex duality} that maps a boson such as a Cooper pair to a dual vortex degree of freedom \cite{Peskin:1977kp,Dasgupta:1981zz,fisher1989}, and plays a key role in our understanding of superfluid-insulator quantum phase transitions. In this letter, we study another particle-vortex duality which transmutes bosonic and fermionic statistics. We demonstrate an exact mapping of partition functions of a strongly coupled boson, and its free fermion vortex.

It is well-known that in $D=2+1$ dimensions, there can be exotic quantum effects such as fractional statistics and statistical transmutation. In a system with a gap, these features can be simply understood as Berry phase effects arising in adiabatic transport \cite{Arovas:1984qr}. Statistical transmutation in a long-wavelength description is implemented via coupling matter to Chern-Simons (CS) gauge fields \cite{Deser1982, Polyakov:1988md, Frohlich1991}. These effective field theories motivate mean-field treatments of the fractional quantum Hall effect based on ``flux attachment'', and have enjoyed much success \cite{zhang1989, Jain:1989tx, lopez1991}. 

A more non-trivial question is whether similar ideas survive in a gapless context, say near a critical point between two quantum Hall phases (see e.g. \cite{Chen:1993cd, Barkeshli:2014ida} for related discussions). Since such systems possess gapless excitations and are in general strongly coupled, the demonstration of statistical transmutation becomes a much more challenging problem. Recently, significant progress towards such a description has been made in the form of a conjectured duality between critical bosons coupled to a level-1 Chern-Simons gauge field and a massless two component Dirac fermion \cite{Aharony:2015mjs, Seiberg:2016gmd,Karch:2016sxi} (see details below). More interestingly, this conjectured boson-fermion duality combined with the familiar boson-vortex duality generate an entire web of dualities \cite{Seiberg:2016gmd, Karch:2016sxi}. Members of this web have important potential applications. Most notably, the fermion-fermion duality \cite{Mross:2015idy} from this web has been proposed to describe the half-filled Landau level \cite{Son:2015xqa} and the surface states of strongly coupled topological insulators \cite{Metlitski:2015eka, Wang:2015qmt, Murugan:2016zal}. On the other hand, from a more formal perspective, along this line people have been exploring non-abelian dualities \cite{Aharony:2015mjs, Hsin:2016blu, Metlitski:2016dht, Aharony:2016jvv, Benini:2017dus} and many more extensions \cite{Kachru:2015rma, Xu:2015lxa, Kachru:2016rui, Karch:2016aux, Mross:2017gny}. It is seen the recent progress on boson-fermion duality has unified many different branches of theoretical physics.

Explicitly, in 3D Euclidean spacetime, the boson-fermion duality maps between the Lagrangian densities
\begin{eqnarray}
 -\mathcal{L}_{E, \, {\rm boson}} &=& -|(\partial_\mu-ib_\mu) \phi|^2 - r|\phi|^2 - \lambda |\phi|^4 + i\frac{\epsilon^{\mu\nu\lambda}}{4\pi} (A-b)_\mu \partial_\nu (A-b)_\lambda \nonumber \\[.2cm]
& \updownarrow& \nonumber \\ -\mathcal{L}_{E, \, {\rm fermion}} &=& \bar\psi \, \sigma^\mu (\partial_\mu - i A_\mu) \, \psi + m\bar\psi\psi + i \frac{\epsilon^{\mu\nu\lambda}}{8\pi} A_\mu \partial_\nu A_\lambda.
\end{eqnarray}
Here $\phi$ is a complex boson, $b$ a dynamical $U(1)$ gauge field, $\psi$ a 2-component Dirac fermion with $\sigma^\mu$ the Pauli matrices, and $A$ a background $U(1)$ gauge field (more precisely, $A$ is a $Spin_c$ gauge field, but this difference does not alter our discussion below). \footnote{We remind the reader that, to avoid gauge ambiguity, the level-1/2 Chern-Simons term on the fermion side must be understood as coming from integrating out a heavy Dirac fermion, with a negative mass whose magnitude is much greater than any physical scale of interest. More formally, one may view this as a Pauli-Villars regularization. In the $m=0$ case, such unambiguously defined level-1/2 Chern-Simons term is usually referred to as $\pi\eta/2$, where $\eta$ is called the $\eta$-invariant \cite{Seiberg:2016gmd}.} This boson-fermion duality holds in the gapped case \cite{Polyakov:1988md} with $\sgn(r)=\sgn(m)$, and, as conjectured, can be extended to the gapless case $r=m=0$. In the gapless case, $\lambda$ is at the Wilson-Fisher fixed point.

While this boson-fermion duality is extremely useful and has undergone many non-trivial consistency checks, it has remained unproven (however, see the demonstrations \cite{Kachru:2015rma, Kachru:2016rui} and \cite{Mross:2017gny}). Recall the familiar boson-vortex duality was originally presented as an essentially exact mapping between the lattice gauge theories of an XY model and an abelian Higgs model \cite{Peskin:1977kp,Dasgupta:1981zz}. A proof of the boson-fermion duality to such rigor is in need.

In this letter, we demonstrate an exact mapping between two 3D Euclidean lattice gauge theories. The first theory is an XY model minimally coupled to a $U(1)$ Chern-Simons gauge field. We implement the Chern-Simons theory on the lattice using gapped Wilson's lattice fermion, thereby manifestly preserving gauge invariance and the compactness of the gauge group. The second theory is a free, massless Dirac fermion, also realized by Wilson's lattice fermion. We show there is an exact mapping relating these two theories which holds even at criticality. Our method is so simple that it can be generalized to other members of the web of dualities \cite{Seiberg:2016gmd, Karch:2016sxi}, and beyond \cite{Aharony:2015mjs, Hsin:2016blu, Metlitski:2016dht, Aharony:2016jvv, Benini:2017dus, Kachru:2015rma, Xu:2015lxa, Kachru:2016rui, Karch:2016aux, Mross:2017gny}. We will develop these ideas elsewhere.

\section{Basic Ingredients}
We consider bosons and fermions on a 3D Euclidean spacetime lattice, taken to be a cubic lattice for simplicity. There are two basic ingredients in the model. The first is a 3D XY model, which has a partition function 
\begin{eqnarray}
&& Z_{XY}[B] = \int D \theta \ e^{- H_{XY}[B]/T}, \ \ \ \ \int D \theta = \prod_n \int_{- \pi}^{\pi} \frac{d\theta_n}{2\pi}, \nonumber \\[.2cm]
&& -H_{XY}[B] = \sum_{n \mu} \cos{\left( \theta_{n + \hat \mu} - \theta_{n} - B_{n \mu} \right)},
\end{eqnarray}
where $n$ labels lattice sites, $\hat \mu = \hat 1, \hat 2, \hat 3$ are elementary displacement vectors, and $n\mu$ labels the link between the sites $n$ and $n+\hat\mu$. The XY model consists of angles $\theta_{n} \in \left( - \pi, \pi \right]$ on each site $n$, and is coupled to a $U(1)$ gauge field $B_{n \mu} \in \left( - \pi, \pi \right]$ on each link $n\mu$. Note that although the lattice XY ``spins'' $\vec s_n = e^{i \theta_n}$ have unit length, upon coarse graining, the amplitude will no longer be constrained. Thus, at distances large compared to the lattice spacing, the XY model describes a complex self-interacting scalar close to its Wilson-Fisher fixed point \cite{Wilson:1971dc}.

The second ingredient will be two sets of Grassmann fields $\bar \chi_n, \chi_n$ describing 2-component Dirac fermions on the lattice sites\footnote{In a Euclidean formulation, Lorentz invariance becomes rotational invariance in $d$ spacetime dimensions. The dimension of the Dirac spinor in this case is $2^{\left[d/2 \right]}$, where $\left[x \right]$ denotes the largest integer $\leq x$.}. We will use Wilson's lattice fermions, which have the following partition function:
\begin{eqnarray}
\label{wilson_fermion}
&& Z_{W} [A] = \int D \bar \chi D \chi \ e^{-H_W[A](M) - H_{\rm int}(U)} , \ \ \ \ \int D \bar \chi D \chi = \prod_n \int d^2 \bar \chi_n d^2 \chi_n, \nonumber \\[.2cm]
% \ \ \ -H_W[A] = \sum_{n n'} \bar \chi_n K_{n n'} \chi_{n'} - H_{int} \nonumber \\
&& -H_W[A](M) = \sum_{n \mu} \left( \bar \chi_n \frac{\sigma^{\mu} - R}{2} e^{-iA_{n \mu}} \chi_{n + \hat \mu} +\bar\chi_{n + \hat\mu} \frac{-\sigma^{\mu} - R}{2} e^{iA_{n \mu}} \chi_n \right) + \sum_n M \bar\chi_n \chi_n. \nonumber \\
%K_{n,n'}[A] &=& \left( \frac{\sigma^{\mu} - R}{2} \right) e^{-iA_{n \mu} } \delta_{n', n + \hat \mu} + \left( \frac{-\sigma^{\mu} - R}{2} \right)e^{iA_{n \mu} } \delta_{n', n - \hat \mu} + M \delta_{n,n'}.
\end{eqnarray}
 $H_{W}$ is the bare theory of massive Dirac fermions on a Euclidean lattice\footnote{We have absorbed the ``temperature'' $T$ into the definition of the fields $\chi_n, \bar \chi_n$.}, with $R$ a constant to be discussed below. The first two terms of the second line of Eq. \eqref{wilson_fermion} are associated with the lattice links whereas the third term, the Dirac mass term, is associated with lattice sites. $H_{{\rm int}}(U)$ is some lattice scale interaction of strength $U$ that will be chosen later (not necessarily physically motivated) for the purpose of constructing a simple exact lattice duality.
%We will choose the interaction
%\begin{eqnarray}
%-H_{\rm int}[A](U) = \sum_{n \mu} \frac{U}{2} \left( \bar \chi_n \frac{\sigma^{\mu} - R}{2} e^{-iA_{n \mu}} \chi_{n + \hat \mu} +\bar\chi_{n + \hat\mu} \frac{-\sigma^{\mu} - R}{2} e^{iA_{n \mu}} \chi_n \right)^2.
%\end{eqnarray}
%Our motivation for this choice will become clear later. 

Suppose we turn off $H_{\rm int}$ for now. Consistent with the doubling theorem \cite{nielsen1981}, there are $2^D = 8$ 2-component Dirac fermion modes in the IR (see appendix). The parameter $R$ was introduced by Wilson as a way to assign different Dirac masses to these Dirac fermion modes \cite{Wilson:1974sk, Wilson1977}. It represents a momentum-dependent correction to fermion masses and is otherwise invisible in the continuum limit. In 3D, even though $R$ gaps out the doubled fermions, these modes never decouple. Instead, they produce Chern-Simons terms. As is well-known, a massive 2-component Dirac fermion with mass $m$ produces a CS term of coefficient $-\sgn(m)/2$. Thus, when all 8 fermion modes are gapped, integrating them out results in the following CS term in the long wavelength limit:
\begin{eqnarray}
-\mathcal{L}_{E, \, {\rm eff}}[A] = i\frac{C}{4 \pi} \epsilon^{\mu \nu \lambda} A_{\mu} \partial_{\nu} A_{\lambda} + (\mbox{terms with higher derivatives}).
\end{eqnarray}
The coefficient $C$ depends on $M, R$ as follows \cite{Golterman:1992ub} (see appendix):
\begin{eqnarray}
C = \left\{ \begin{array}{ll}
0, & \ \ \ 3|R|<|M| \\
\sgn(R), & \ \ \ |R|<|M|<3|R| \\
-2\sgn(R), & \ \ \ |M|<|R|. \\
\end{array} \right.
\label{C_in_MR}
\end{eqnarray}
%For technical convenience below, we shall take $R = +1$.
We will address the effects of $H_{\rm int}$ later.

The partition function for a boson coupled to a CS term can now be written:
\begin{eqnarray}
Z[A] = \int Db \ Z_{XY}[b] \ Z_{W}[A-b], \ \ \ \ \int Db = \prod_{n \mu} \int_{-\pi}^{\pi} \frac{db_{n \mu}}{2\pi}
\end{eqnarray}
where $b_{n \mu}$ is a fluctuating $U(1)$ gauge field that lives on the links. We have not imposed a gauge fixing condition for $\int Db$, since the gauge redundancy is just an overall factor. Provided that $H_{W}+H_{\rm int}$ satisfies certain UV boundary conditions to be described below, integrating out the fermions above will lead to a level-1 CS term for the difference $(A-b)_{\mu}$. Since the XY boson couples to $b_{\mu}$, we expect this will impart statistical transmutation and convert the boson into a fermion.

\section{Exact UV Mapping}
We next present an exact mapping from the partition function $Z[A]$ above to that of a Wilson fermion with a different mass $M'$. To do this, we work in the representation of boson currents $j_{n \mu}$ defined on the links, obtained by Fourier series \cite{Frohlich:1981yn}:
\begin{eqnarray}
e^{\frac{1}{T} \cos{
\left( \theta_{n} - \theta_{n + \hat \mu} - b_{n \mu} \right)}} = 
\sum_{j_{n \mu} =-\infty}^{\infty} 
I_{j_{n \mu}}(1/T) \:
e^{i \left( \theta_n - \theta_{n + \hat \mu} - b_{n \mu} \right) j_{n \mu}}
\end{eqnarray}
where $I_j(x)=I_{-j}(x)$ is the $j$th modified Bessel function. It represents a sum over all possible boson tunneling terms, each term resulting in moving an integer $j_{n \mu}$ number of bosons across the link, with a tunneling amplitude $I_{j_{n \mu}}(1/T)$. Integrating out $\theta_n$ leads to a ``Gauss's Law'' constraint on each site:
\begin{eqnarray}
Z_{XY}[b] =  \sum_{ \{ j_{n \mu} \} } \left( \prod_{n \mu} I_{j_{n \mu}}(1/T) \: e^{-i b_{n \mu} j_{n \mu} } \right) \left( \prod_{n} \delta_{\Delta_{\mu} j_{n \mu}} \right), \ \ \ \   \sum_{ \{ j_{n \mu} \} } = \prod_{n \mu} \sum_{j_{n \mu} = -\infty}^{\infty}.
\end{eqnarray}
The divergence-free condition is of course just a restatement of the $U(1)$ conservation of the original XY model. %In continuum language, emergent gauge fields arise since the conserved current can be expressed in terms of a dual photon: $j_{\mu} \propto \epsilon^{\mu \nu \lambda} \partial_{\nu} a_{\lambda}$. Instead of following this path, which would lead to the dual Abelian Higgs theory of bosonic vortices, 

\begin{figure}[t]
\centering
\includegraphics[width=3.2in]{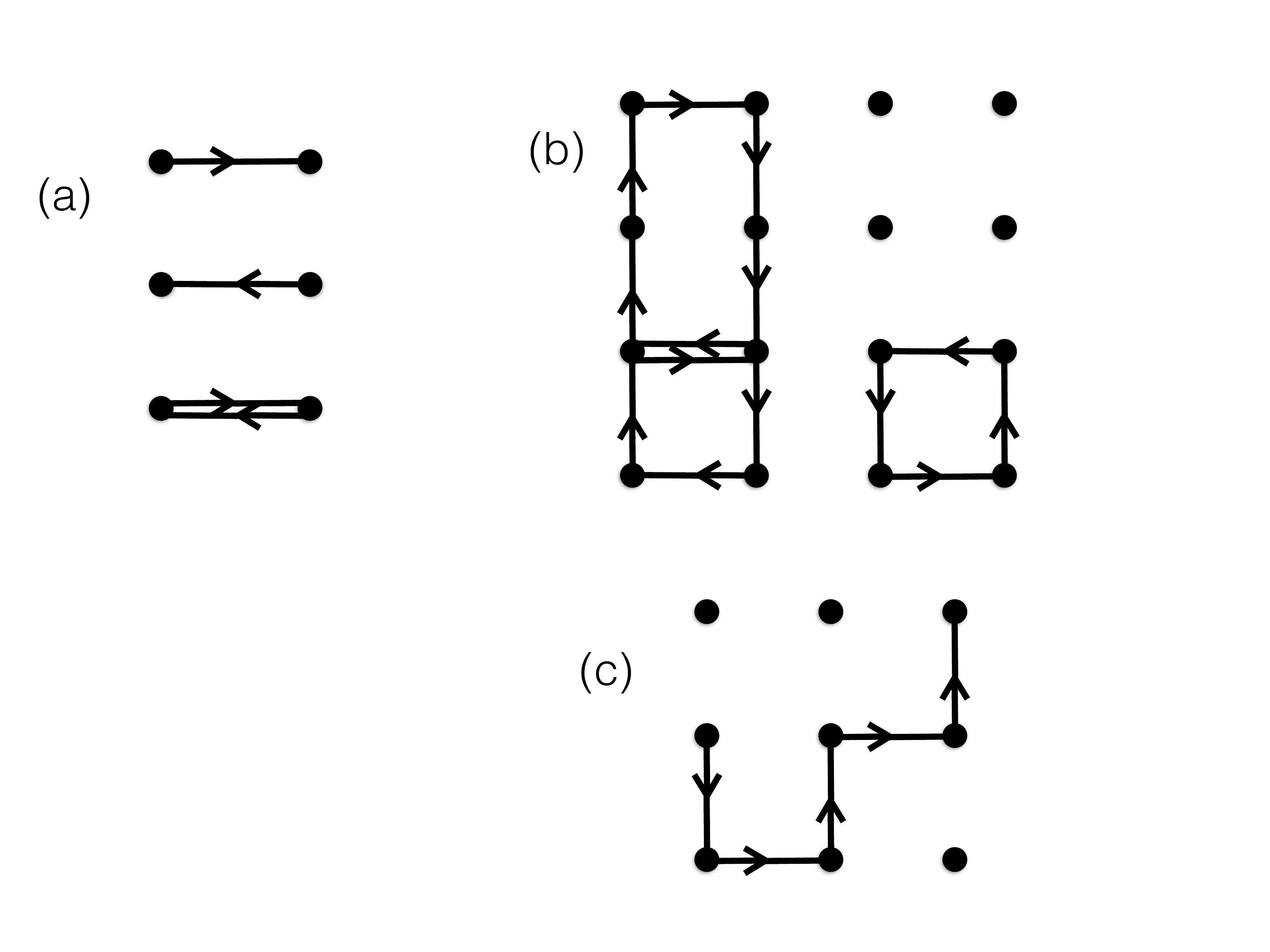}
\caption{\small (a) The various terms that arise on a link $n \mu$ in the exact expansion of Grassmann fields in $Z_W$. From top to bottom, the contributions are: hopping $\bar \chi_{n + \hat \mu} \frac{-\sigma^{\mu} - 1}{2} \chi_n$, hopping $ \bar \chi_n \frac{\sigma^{\mu} -1}{2} \chi_{n + \hat \mu}$, double hopping plus interaction $(1+U) \left( \bar\chi_n \frac{\sigma^{\mu}-1}{2} \chi_{n+\hat{\mu}} \right) \left( \bar\chi_{n+\hat{\mu}} \frac{-\sigma^{\mu}-1}{2} \chi_n \right)$. (b) In a Grassmann integral, each fermion component must appear exactly once. Consider a conjugate pair of fermion components, say $\chi_{n \uparrow}$ and $\bar\chi_{n \uparrow}$. They either appear together in a mass term, or appear separately in two link terms. So the link terms always form closed loops. If this condition is not satisfied as in (c), the contribution vanishes by Grassmann algebra. Thus, all contributions to $Z_W$ manifestly satisfy Gauss's law. (The lattice is 3D. We drew a 2D lattice for clarity.)}
\label{lattice_fermion_diagrams}
\end{figure}

We then consider the Wilson fermions. Following Wilson, we set $R=+1$. The choice of the magnitude $1$ does not affect the IR physics, but it simplifies the lattice model as $\pm \sigma^\mu - 1$ project out one linear combination of the 2-component fermion. For reasons to be made apparent below, for $|R|=1$ we choose $H_{\rm int}$ to be a hopping-hopping interaction:
\begin{eqnarray}
-H_{\rm int}(U) &=& \sum_{n \mu} \frac{U}{2} \left( \bar \chi_n \frac{\sigma^{\mu} - R}{2} e^{-iA_{n \mu}} \chi_{n + \hat \mu} +\bar\chi_{n + \hat\mu} \frac{-\sigma^{\mu} - R}{2} e^{iA_{n \mu}} \chi_n \right)^2 \nonumber \\[.2cm]
&=& \sum_{n \mu} U \left( \bar\chi_n \frac{\sigma^{\mu}-R}{2} \chi_{n+\hat{\mu}} \right) \left( \bar\chi_{n+\hat{\mu}} \frac{-\sigma^{\mu}-R}{2} \chi_n \right).
\end{eqnarray}
Due to the Grassmann algebra of the fermion fields, we may expand the contribution to $Z[A]$ on each link \emph{exactly} (see Fig. \ref{lattice_fermion_diagrams}). Each link $n \mu$ contributes a factor $Z_{n \mu}$ which is
\begin{eqnarray}
%&&Z[A]=\prod_{n \mu} Z_{n \mu}[A], 
Z[A] &=& \int D \bar \chi D \chi \ \sum_{ \{ j_{n \mu} \}}  \ \left( \prod_{n} \delta_{\Delta_{\mu} j_{n \mu}} e^{M \bar \chi_n \chi_n} \right) \left(\prod_{n \mu} Z_{n \mu} \right), \nonumber \\[.2cm]
Z_{n \mu} &=& \int_{-\pi}^\pi \frac{db_{n \mu }}{2\pi} \  I_{j_{n \mu }}(1/T) \ e^{-ib_{n \mu } j_{n \mu}} \nonumber \\[.1cm]
&& \left[ 1 + e^{-i(A-b)_{n \mu}}\bar\chi_n \frac{\sigma^{\mu}-R}{2} \chi_{n+\hat{\mu}} + e^{i(A-b)_{n \mu}} \bar\chi_{n+\hat{\mu}} \frac{-\sigma^{\mu}-R}{2} \chi_n \right. \nonumber \\[.1cm]
&& \ \ \ \, \left. + \ (1+U) \left( \bar\chi_n \frac{\sigma^{\mu}-R}{2} \chi_{n+\hat{\mu}} \right) \left( \bar\chi_{n+\hat{\mu}} \frac{-\sigma^{\mu}-R}{2} \chi_n \right) \right].
\end{eqnarray}
The remarkable feature of the expression above is that the integration over the dynamical gauge field $b_{n \mu}$ can be performed \emph{exactly}. Doing so, 
%we find that boson tunneling operators are vastly constrained: only $j_{n \mu} = 0, \pm 1$ bosons can now tunnel across each lattice link\footnote{The constraint $\Delta_{\mu} j_{n \mu} = 0$ is not violated in this step. To see this explicitly, we may define $j_{n \mu} = \epsilon^{\mu \nu \lambda} \Delta_{\nu} a_{\tilde n \lambda}$ where $a_{\tilde n \mu}$ live on the links of the dual lattice. Integration over $b_{n \mu}$ leads to the constraint that the fluxes $\vec \Delta \times \vec a$ on each of the four plaquettes associated with a link $n \mu$ satisfy $\epsilon^{\mu \nu \lambda} \Delta_{\nu} a_{\tilde n \lambda} = 0, \pm 1$}. This behavior is exactly that of fermion fields.
we find the boson tunneling current $j_{n\mu}$ must be equal to the fermion tunneling current, which can only take values $0, \pm 1$ as seen in the Grassmann expansion above. This shows the bosons and the fermions always form composite particles. Note that the Gauss's Law constraint $\Delta_{\mu} j_{n \mu} = 0$ for bosons is \emph{automatically} satisfied, because under the Grassmann integral, fermion tunneling currents must form closed loops to have non-vanishing amplitude, as shown in Fig. \ref{lattice_fermion_diagrams}. 
%\footnote{Here our $\int Db$ has an overall factor of gauge redundancy. Suppose we instead impose the axial gauge $b_{n3}=0$. Then integrating out $b_{n\mu}$ only bounds $j_{n\mu}$ to fermion tunneling currents for $\mu=1, 2$. However, since both fermion tunneling currents and $j_{n\mu}$ satisfy the Gauss's Law, the bounding must also hold for $\mu=3$. This always works out because the Gauss's Law constraint is related to gauge invariance.}
Thus, after integrating out $b_{n\mu}$,
\begin{eqnarray}
\sum_{j_{n \mu} = -\infty} ^ {\infty}Z_{n \mu} &=& I_{0}(1/T)
 \left[ 1 + \frac{I_1(1/T)}{I_0(1/T)} e^{-iA_{n \mu}} \bar\chi_n \frac{\sigma^{\mu}-R}{2} \chi_{n+\hat{\mu}} + \frac{I_{1}(1/T)}{I_0(1/T)} e^{iA_{n \mu}} \bar\chi_{n+\hat{\mu}} \frac{-\sigma^{\mu}-R}{2} \chi_n \right. \nonumber \\[.1cm]
&& \hspace{1.9cm} \left. + \ (1+U) \left( \bar\chi_n \frac{\sigma^{\mu}-R}{2} \chi_{n+\hat{\mu}} \right) \left( \bar\chi_{n+\hat{\mu}} \frac{-\sigma^{\mu}-R}{2} \chi_n \right) \right],
\end{eqnarray}
which describes fermions with renormalized tunneling amplitudes and self-interactions on each link. It is now clear why we have chosen $H_{\rm int}$ to be the hopping-hopping interaction. Redefining the fermion fields $\psi_n = \sqrt{I_1(1/T)/I_0(1/T)} \: \chi_n$ (and likewise for $\bar\psi_n$), we can see a hopping-hopping interaction is generated from the coupling to bosons, so we want to include a bare $H_{\rm int}$ that has the same form as the generated interaction, to act as a ``counterterm'' and ultimately to produce a free fermion in the IR (see next section). More explicitly,
%We view $H_{\rm int}$ as a set of lattice scale interactions that can be adjusted as needed to produce the desired long-wavelength physics. With the expression above, it becomes clear how to choose the interactions: since we wish to obtain a free fermion from the duality, we can choose the interactions in $H_{\rm int}$ to cancel the interactions generated by the exact expansion of Grassmann fields. Thus, we choose the hopping-hopping interaction
%\begin{eqnarray}
%H_{\rm int} = \sum_{n \mu} H_{{\rm int}, n \mu} , \ \ H_{{\rm int},n \mu} = - \frac{U}{2} \left( \bar \chi_n \frac{\sigma^{\mu} - R }{2} \chi_{n + \hat \mu} + \bar \chi_{n + \hat \mu} \frac{ -\sigma^{\mu} - R}{2} \chi_n \right)^2.
%\end{eqnarray}
%Redefining fermion fields $\psi_n = \sqrt{I_1(1/T)/I_0(1/T)} \chi_n$,
the total partition function of the system in terms of $\psi, \bar\psi$ reads
\begin{eqnarray}
Z[A] \propto \int D \bar \psi D \psi \ e^{-H_W[A](M') - H_{\rm int}(U')}, \ \ \ \ \ \frac{M'}{M} = \frac{I_0(1/T)}{I_1(1/T)} = \sqrt{ \frac{1+U'}{1+U} }
%-H'_W &=& \sum_{n \mu}\left( \bar \psi_n \frac{\sigma^{\mu} - R}{2} e^{-i A_{n \mu} }\psi_{n + \hat \mu} + \bar \psi_{n+ \hat \mu} \frac{-\sigma^{\mu} - R}{2} e^{i A_{n \mu} }\psi_{n } \right) + \sum_n M' \bar \psi_n \psi_n,
\label{renorm}
\end{eqnarray}
which describes a Wilson fermion with modified $M'$ and $U'$; the function ${I_0(1/T)}/{I_1(1/T)}$ increases from $1$ to $+\infty$ as $T$ increases from $0$ to $+\infty$. This is the exact lattice duality we claimed.

By similar steps, it is also easy to verify that the correlation functions satisfy
\begin{eqnarray}
\left\langle e^{i\theta_{n_1}}\chi_{n_1} \cdots e^{i\theta_{n_k}}\chi_{n_k} \, e^{-i\theta_{n'_1}}\bar\chi_{n'_1} \cdots e^{-i\theta_{n'_k}}\bar\chi_{n'_k} \right\rangle_{[A]} \propto \left\langle \psi_{n_1} \cdots \psi_{n_k} \bar\psi_{n'_1} \cdots \bar\psi_{n'_k} \right\rangle_{[A]}
\end{eqnarray}
under any background $A_{n\mu}$ configuration. This again manifests the point that the fermion $\psi$ is the composite particle formed by the boson $e^{i\theta}$ and the fermion $\chi$ due to the mediation of the gauge field $b$. Note that $e^{i\theta} \chi$ is invariant under the $U(1)_b$ gauge transformation of $b$, as $\psi$ should be.

\section{UV Boundary Conditions}
\begin{figure}[t]
\centering
\includegraphics[width=3in]{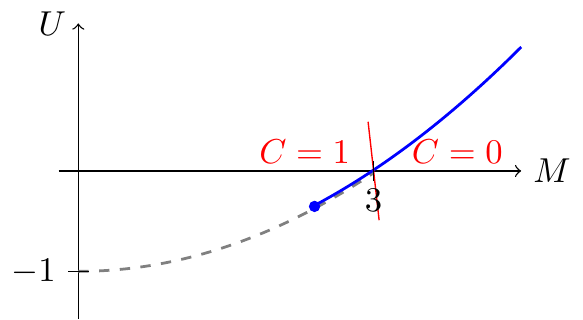}
\caption{\small The gray dashed curve (section of parabola) contains the values of $M, U$ that satisfy Eq. \eqref{renorm1} for some $T_c$. Suppose we have chosen some $M, U$ on the gray dashed curve, say the blue dot. Then as $T$ increases from $0$ to $+\infty$, the corresponding $M', U'$ trace up along the blue section of the parabola. The red line represents the phase boundary between the $C=1$ and $C=0$ phase regions in the vicinity of $M=3, U=0$. The slope is $\approx -8.8$ but we did not compute higher derivatives, so we do not know how the phase boundary looks further away from the point $M=3, U=0$. However, this is sufficient to show our choice of blue dot lies within the $C=1$ phase as desired.}
\label{phase_diagram}
\end{figure}

In the previous section, we have shown that the 3D XY model coupled to a massive interacting Wilson fermion $\chi$ can be exactly mapped in the UV to another (generically massive and interacting) Wilson fermion $\psi$. In order to realize the conjectured IR duality, we will need to adjust the parameters of the UV theory, so that $\chi$ is massive and can be integrated out before the gauge field $b$ to implement a level-1 CS, \footnote{This can be viewed as \emph{defining} what the strongly interacting ``boson+CS'' Lagrangian in continuum actually means, since a continuum Lagrangian is generally not meaningful without proper regularization conditions.} while $\psi$ has a free massless Dirac fermion mode in the IR. For concreteness, to realize the IR behavior of $\psi$, we can set $M'=3, U'=0$, so that the Dirac mode near lattice momentum $k_\mu = 0$ becomes massless, while the other 7 Dirac modes contribute a level-1/2 CS for the background field $A$ (see appendix). From Eq. \eqref{renorm}, we see that $M'=3, U'=0$ are realized when the UV parameters $M,U$ of $\chi$ satisfy 
\begin{eqnarray}
1+U = \frac{M^2}{9} = \left( \frac{I_1(1/T_c)}{I_0(1/T_c)} \right)^2
\label{renorm1}
\end{eqnarray}
where $T_c$ is the value of the ``temperature'' $T$ of the XY model required to hit the critical point. Such values of $M$ and $U$ lie on a section of a parabola shown by the gray dashed curve in Fig. \ref{phase_diagram}. However, now $U$ \emph{must be} non-zero. We know that when $U=0$, a Wilson fermion with $1<M<3$ can be integrated out to generate a CS term of level $C=1$. Now we need to show $C=1$ continues to hold when $M, U$ lie on the said section of the parabola, at least in the vicinity $M\lesssim 3, U\lesssim 0$. This way, the heavy Wilson fermion $\chi$ continues to implement the level-1 CS term for $(A-b)$ in the IR.

Consider the phase diagram in $M$ and $U$. Massive fermions with $C=1$ and $C=0$ are in different topological phases, so there must be a gapless phase boundary separating the two phase regions, and we know this phase boundary goes through the point $M=3, U=0$. We claim the phase boundary looks like the red line in Fig. \ref{phase_diagram} in the vicinity of the point $M=3, U=0$. To justify this claim, it is sufficient to consider $M=3+\delta M, U=0+\delta U$ and determine the slope $\delta U/\delta M$ such that the Dirac mode near $k_\mu=0$ remains massless. That is, we equate $\delta M + \Sigma = 0$, where $\Sigma$ is the self-energy evaluated at $k_\mu=0, M=3$ to first order in $\delta U$. The slope thus computed is $\delta U/\delta M \approx -8.8$, as drawn in Fig. \ref{phase_diagram}. Therefore, at least in the vicinity $M\lesssim 3, U\lesssim 0$, points satisfying Eq. \eqref{renorm1} lie in the $C=1$ phase as desired. This completes the verification that the constructed UV duality reproduces the conjectured IR duality. One can also check that when $T$ deviates from $T_c$, the IR mass of $\psi$ has the expected sign.

%Since the hopping-hopping interaction $H_{\rm int}(U')$ is itself irrelevant in the IR, the only effect of $U'$ is its renormalization of the IR mass. Therefore we could have chosen $M', U'$ to be other points on the gapless phase boundary. Our particular choice of $M'=3, U'=0$ is a matter convenience rather than fine tuning.

%{\it I would propose the paragraph below as an alternative to the one above, though I'm not exactly sure what was being conveyed above, so the paragraph below may be different -- SR.} \\
The reader may naturally wonder whether our analysis of the UV boundary condition amounts to fine-tuning beyond that which is required to reach a critical point. However, since $H_{\rm int}(U')$ is irrelevant near the free IR fixed point, its only effect is to renormalize the IR mass. Put differently, $M', U'$ are not independent parameters that need to be tuned (although we tuned to $M'=3, U'=0$ for demonstration). \footnote{In the language of effective field theory, both $M$, and $U$ are renormalized by fermion wavefunction renormalization $Z_{\psi}(T)$, which in turn only depends on $T$. Thus, only one parameter needs to be tuned to access the critical point as is always the case.} The irrelevancy of the interaction in the IR also guarantees that first-order transitions are unlikely for small deviations from the point $M'=3, U'=0$. For large deviations from this point, a perturbative analysis is not controlled and other possibilities may occur. 

On the other hand, the behavior of the phase boundary(ies) between the $C=-2$ and $C=1$ phases in the vicinity of $M=1, U=0$ remains unclear.

\section{Conclusion}

In this letter, we solved a strongly interacting 3D lattice gauge theory exactly and showed that the solution reproduces a conjectured boson-fermion duality in the long distance limit.

%We have constructed an exact UV duality on a 3D Euclidean lattice, and shown it reproduces the conjectured boson-fermion duality in the IR.
Our idea of realizing $\psi$ as the composite particle $e^{i\theta} \chi$ is not new -- it resembles the so-called ``parton construction'' widely used in theories of fractional quantum Hall systems \cite{Jain:1989, Blok:1990mc}. 
%However, the usual parton construction in the continuum is not mathematically well-defined, and the partons are at best heuristic constructions whose justification can only be made {\it a posteriori}.
%On the other hand,
In our lattice construction, the following facts are all derived exactly: 1) the boson $e^{i\theta}$ and the fermion $\chi$ form a composite particle $\psi$, 2) the fermion $\chi$ is massive and can be integrated out to produce level-1 Chern-Simons, and 3) the fermionic composite particle $\psi$ becomes massless. In the IR duality, the fermion $\psi$ is viewed as the composite particle $\phi \mathcal{M}$ where $\mathcal{M}$ is the so-called ``monopole operator''. Clearly $\phi$ has been realized as the coarse-grained version of $e^{i\theta}$. Moreover, $\mathcal{M}$ can indeed be realized by the heavy fermion $\chi$, because a heavy fermion excitation moving in a Dirac sea in the $C=1$ phase attains a $2\pi$ flux on itself.

One may ask to what extent our ``XY + heavy fermion'' lattice construction represents the ``Wilson-Fisher + CS'' side of continuum duality. In a sense, our construction can be viewed as \emph{giving} an operational definition to the strongly coupled  continuum Lagrangian by providing a non-perturbative regularization. Then our claim is that, after integrating out the heavy fermion $\chi$ and the fast modes of bosonic fields $e^{i\theta}$ and $b$, the slow modes of the bosonic fields ``look sufficiently like'' the continuum Lagrangian of Wilson-Fisher + CS. Claims of such kind, generally not justifiable by analytic means, are implicitly understood in common applications of lattice gauge theory such as lattice QCD \cite{Wilson:1974sk, Wilson1977} and the original presentation of the boson-vortex duality \cite{Peskin:1977kp,Dasgupta:1981zz}. Our procedure is a derivation of the duality within such criterion.

%{\it We need to have a final paragraph that reinforces future work.  Here is a draft of one --SR} \\
%The exact mapping of lattice gauge theories presented here may be extended to other members of the web of dualities. The non-perturbative lattice regularizations of these theories enables the study of non-relativistic dualities, often applicable in condensed matter contexts. Our methods can also be used to study dualities involving matter fields with non-abelian gauge fields. We will report on progress in these directions in forthcoming publications. 
%{\it I know it's repetitive, but unless there is something else to say it is worth mentioning this point again.  -- SR.} \\

\

\noindent \emph{Acknowledgments - } 

We thank Chao-Ming Jian, Nathan Seiberg, Dam Thanh Son and Xiao-Qi Sun for discussions, and thank Shamit Kachru and David Mross for comments on the manuscript. J.-Y.~C. is supported by the Gordon and Betty Moore Foundation's EPiQS Initiative through Grant GBMF4302. S.~R. is supported by the DOE Office of Basic Energy Sciences, contract DE-AC02-76SF00515.

\appendix
\section{Chern-Simons Term from Wilson Fermions}
Consider a non-interacting Wilson fermion with all gauge fields switched off. Then we can transform to momentum space where $\chi_n = \int_{-\pi}^\pi \frac{d^3 k}{(2\pi)^3} e^{ik\cdot n} \chi_k$, $\bar\chi_n = \int_{-\pi}^\pi \frac{d^3 k}{(2\pi)^3} e^{-ik\cdot n} \bar\chi_k$, so that
\begin{align}
-H_W[A=0] = \int_{-\pi}^\pi \frac{d^3 k}{(2\pi)^3} \ \bar\chi_k \left( \sum_\mu\left(\sigma^\mu\: i\sin k_\mu -R \cos k_\mu \right) +M \right) \chi_k.
\end{align}
For each $k$ there are two eigenvalues $M-\sum_\mu R\cos k_\mu \pm i\left(\sum_\mu \sin^2 k_\mu\right)^{1/2}$, whose product has 8 extrema in the Brillouin zone, located at where each $k_\mu$ component takes value either $0$ or $\pi$, corresponding to 8 Euclidean Dirac modes in the continuum. If all $k_\mu$ components are $0$, the mass is $+(M-3R)$; if one is $\pi$ and two are $0$, the mass is $-(M-R)$; if two are $\pi$ and one is $0$, the mass is $+(M+R)$; if all are $\pi$, the mass is $-(M+3R)$. Each mode (assuming non-zero mass) contributes $-\sgn(mass)/2$ to $C$, leading to Eq. \eqref{C_in_MR}. The same result can also be directly computed from the 1-loop current-current correlation \cite{coste1989parity}.

\bibliography{Bose-Fermi_duality}{}
\bibliographystyle{utphys}
\end{document}